\documentclass[12pt]{article}
\usepackage{esfconf}

\newcommand{\bhg}{\beta_{\rm hg}}
\newcommand{\thg}{T_{\rm hg}}
\newcommand{\Ord}{{\cal{O}}}
\renewcommand{\sp}{\ ,\qquad}

\newcommand{\Zint}{Z}
\newcommand{\eqref}[1]{(\ref{#1})}
\renewcommand{\sp}{\ ,\qquad}
\newcommand{\nn}{\nonumber}

\newcommand{\leff}{l_{\rm eff}}
\newcommand{\spa}{\ \ ,\ \ }
\newcommand{\gym}{g_{\rm YM}}

\begin{document}


\hfill NBI-HE-00-35 \\
${}$ \hfill SPIN-2000/25 \\
${}$ \hfill ITF-UU-00/27 \\ 
${}$ \hfill hep-th/0010169 \\
 
\vskip 2cm 
\title{Hagedorn Behavior of Little String Theories\footnote{\small
Talk presented by N.O. at conference "Quantization, Gauge Theory, and
Strings'', dedicated to the memory of Prof. E. Fradkin, Moscow,
Russia, June 5-10.}}


\authors{T. Harmark\adref{1} and N.A. Obers\adref{2,3}}


\addresses{\1ad
Niels Bohr Institute, Blegdamsvej 17, DK-2100 Copenhagen, Denmark,
\nextaddress \2ad
Spinoza Institute, Utrecht University, 3584 CE Utrecht, The
Netherlands
\nextaddress
\3ad
Institute for Theoretical Physics, Utrecht Universty \\
3508 TA Utrecht, The Netherlands
}


\maketitle


\begin{abstract}
We examine the Hagedorn behavior of little string theory using its
 conjectured duality with near-horizon NS5-branes. In particular, by
 studying the string-corrected NS5-brane supergravity solution, it
 is shown that tree-level corrections to the temperature vanish,
 while the leading one-loop string correction generates the
 correct temperature dependence of the entropy near the Hagedorn
 temperature. Finally, the Hagedorn behavior of OD$p$-brane theories, which
are deformed versions of little string theory, is considered via
 their supergravity duals.
\end{abstract}


\newpage 
\section{Introduction}
The near-horizon limit of NS5-branes is conjectured to be dual
to Little String Theory (LST) with 16 supercharges \cite{Aharony:1998ub}.
LST is a 5+1 dimensional non-gravitational and
non-local theory of strings
\cite{Seiberg:1997zk} (see also  
\cite{Dijkgraaf:1997ku}). As for any
string theory, the statistical mechanics description of LST breaks
down at a certain temperature, known as the Hagedorn temperature
\cite{Hagedorn:1965st} (see also e.g.  
\cite{Salomonson:1986eq}).  
This raises the question whether it is possible to observe, via
the conjectured near-horizon-NS5/LST duality, Hagedorn behavior
of LST from the thermodynamics of near-horizon NS5-branes.

Indeed, in the work of
\cite{Maldacena:1996ya,Maldacena:1997cg} it was shown that
the leading order thermodynamics of non-extremal near-horizon
NS5-branes corresponds to the leading order Hagedorn
behavior, since one finds the relation $ E = T S $ where $T$ is
constant. However, since the energy dependence of the entropy
near the Hagedorn temperature for a (six-dimensional supersymmetric)
string theory is of the form 
\begin{equation}
\label{entr} S(E) =\bhg E + k \log E \sp  \bhg = \thg^{-1} 
\end{equation}
this apparently only reproduces the first term in the entropy expression.
This raises the question of how the NS5-brane background can reproduce
\eqref{entr}, or equivalently
\begin{equation}
\label{entr0} S(T) = k \frac{\thg}{\thg- T} \ , 
\end{equation}
which exhibits the near Hagedorn behavior of the entropy as a function
of the temperature.
Another related issue is that the
leading order NS5-brane thermodynamics has
a degenerate phase space. In this talk, we will show that
the resolution to these problems is to
incorporate string corrections
to the NS5-brane supergravity solution \cite{Harmark:2000hw} (see also
\cite{Berkooz:2000mz}). In particular, we reproduce the temperature 
dependence of \eqref{entr0}, where $k$ is a constant determined by
the one-loop corrected near-horizon NS5-brane background. 
This is therefore an example of agreement between statistical thermodynamics
of a non-gravitational theory on the one hand and Bekenstein-Hawking
thermodynamics of a black brane on the other hand.
We note that we can compute the thermodynamics from both sides
since we are in a regime in which both
LST and the bulk string theory description are weakly coupled.
This successful comparison lends further confidence to the conjecture
that the various dualities between string and M-theory on
near-horizon brane backgrounds
and certain non-gravitational theories hold
beyond the leading order supergravity solution.

The plan and summary of the talk is as follows. We first review
the definition of LST and
discuss the dual supergravity solution and its thermodynamics,
followed by an analysis of the tree-level and one-loop string
corrections arising from the $R^4$ term in the type II action. We
will then show that the latter corrections indeed reproduce the
expected temperature dependence \eqref{entr0} of the entropy.
Finally, we will comment on the thermodynamics of
a number of related non-gravitational theories in six dimensions,
known as OD$p$-brane theories  \cite{Gopakumar:2000ep,Harmark:2000ff},
that have recently have been found,
along with their supergravity duals.
For these theories we derive the thermodynamics as well, showing
a different Hagedorn behavior, with critical exponent $-\frac{2}{3}$,  
 the explanation of which is still an open problem.

\section{Hagedorn behavior of LST from string corrections to NS5-branes}

Six-dimensional LSTs with 16 supercharges can be defined
from $N$ coincident NS5-branes
in the limit \cite{Seiberg:1997zk}
\begin{equation}
\label{decoup} g_s \rightarrow 0 \sp l_s = \mbox{fixed} \ ,
\end{equation}
with $g_s$ and $l_s$ being respectively the string coupling
and string length of type II string theory.
In the limit \eqref{decoup} the bulk modes decouple and
give rise to non-gravitational theories: The $(1,1)$ LST for type IIB
NS5-branes and the $(2,0)$ LST for type IIA NS5-branes, both
of type  $A_{N-1}$. 
Here, ,
we will use the conjecture that string theory in the background of
$N$ NS5-branes is dual to LST \cite{Aharony:1998ub}, in parallel
with the AdS/CFT correspondence \cite{Maldacena:1997re}. In order
to be at finite temperature the correspondence for the
near-extremal rather than the extremal case will be considered here.
We start by considering the supergravity solution of $N$ coincident
NS5-branes in type II string theory with $r$ the transverse radius
and $r_0$ the horizon radius
\begin{equation}
  ds^2
= H^{-1/4} \left[ \Big( 1 - \frac{r_0^2}{r^2} \Big)  dt^2 +
\sum_{i=1}^5 (dy^i)^2  + H \Big( \Big( 1 - \frac{r_0^2}{r^2}
\Big)^{-1}
 dr^2 + r^2  d \Omega_{3}^2 \Big) \right] \ , 
\end{equation}
\begin{equation}
 e^\phi = g_s H^{1/2} \qquad , \qquad  H = 1 + \frac{N l_s^2}{r^2} \ . 
\end{equation}
Then, if we take the decoupling
limit \eqref{decoup} keeping fixed the energy scales
\begin{equation}
\label{ufixed}
u = \frac{r}{g_s l_s^2} \sp
u_0 = \frac{r_0}{g_s l_s^2}
\end{equation}
one obtains the Einstein-frame metric and dilaton
\begin{equation}
\label{NS5met}
\frac{ds^2}{\sqrt{g_s} l_s}
 = \sqrt{\frac{u}{\sqrt{N} l_s }}
 \left[ - \left(1-\frac{u_0^2}{u^{2}} \right) dt^2
+ \sum_{i=1}^5 (dy^i)^2 + N l_s^2 \left( \frac{du^2}{u^2-u_0^2} +
d \Omega_{3}^2 \right) \right] \ ,
\end{equation}
\begin{equation}
\label{NS5dil} g_s e^\phi = \frac{\sqrt{N}}{l_s u} \ .
\end{equation}
This supergravity solution is conjectured to be dual to a LST
\cite{Aharony:1998ub}.
The variable $u$ defined by \eqref{ufixed} is kept finite
since in type IIB it corresponds  to the mass
of an open D-string stretching between two NS5-branes with
distance $r$. In type IIA, $u/l_s$  is instead the induced
string tension of an open D2-brane
stretching between two NS5-branes.

The curvature $e^{-\phi/2}R$ in units of $l_s^{-2}$ and
the effective string coupling squared at the horizon radius
are respectively of order
\begin{equation}
\label{epsD} \varepsilon_D = \frac{1}{N} \sp
 \varepsilon_L =  \frac{N}{l_s^2 u_0^2} \ .
\end{equation}
In order for the near-horizon NS5-brane solution \eqref{NS5met},
\eqref{NS5dil} to be a dual description of LST we need that
\( \varepsilon_D \ll 1 \) and $\varepsilon_L \ll 1$
 which is fulfilled for \cite{Maldacena:1997cg,Aharony:1998ub}
\begin{equation}
\label{limits} N \gg 1 \sp l_s^2 u_0^2 \gg N \ ,
\end{equation}
so that near-horizon NS5-branes describe LST in the UV-region.
The thermodynamics%
\footnote{The thermodynamics of spinning near-horizon NS5-branes is
found in \cite{Sfetsos:1999pq,Harmark:1999xt}.}
of the near-horizon NS5-brane solution is
\cite{Maldacena:1997cg}
\begin{equation}
\label{leadtherm}
T = \frac{1}{2\pi \sqrt{N} l_s} \sp
S = \frac{\sqrt{N} V_5}{(2\pi)^4 l_s^3 } u_0^2 \sp
E = \frac{V_5}{(2\pi)^5 l_s^4} u_0^2 \sp
F = 0 \ .
\end{equation}
Indeed, this is  \cite{Maldacena:1996ya,Maldacena:1997cg},
the zeroth order approximation
the thermodynamics $S = \bhg E$ of a string theory at high temperature
by identifying the Hagedorn temperature as
\begin{equation}
\label{thag}
\thg = (2\pi \sqrt{N} l_s )^{-1}
\end{equation}
On the other hand, the Hagedorn temperature of a closed
string theory with string length $\hat l_s$ and central charge
$c$ is given by $ \thg = (2\pi \hat l_s \sqrt{6/c} )^{-1}$
so that for a 5+1 dimensional supersymmetric string theory
with $c=6$, one has $\thg = (2\pi
\hat{l}_s)^{-1}$.
Comparing this with \eqref{thag}, on observes that
$ \hat{l}_s = \sqrt{N} l_s$ 
is the string length associated with the Hagedorn exponential growth of
string states in LST of type $A_{N-1}$.
In terms of the string tension, $\hat{\tau} = \tau/N$
this shows that the string tension associated with the Hagedorn behavior
is quantized in a unit that is a fraction of
the ordinary string tension. Indeed, by considering  
LST as the decoupling limit \eqref{decoup}
of type II string theory on an $A_{N-1}$ singularity \cite{Ooguri:1996wj},
it is known from the study of string theory on this orbifold singularity
\cite{Douglas:1996sw} that there exist
fractional ($1/N$) branes. 
Another way to argue that the Hagedorn temperature picks the
$1/\hat{l_s}$
string scale is that this is the first one reached when raising the
temperature \cite{Aharony:1998ub}. 

We now want to include string corrections in the thermodynamics%
\footnote{We refer to \cite{Correia:2000} for computational
details and a general analysis of string corrections to near-horizon
brane thermodynamics.}, arising from the higher derivative terms
and string loops.  
It follows from \eqref{epsD} that the derivative expansion,
which is an expansion in $\alpha'=l_s^2$,
becomes an expansion in $\varepsilon_D$ for the
near-horizon NS5-brane solution.
The tree-level expansion of the temperature is
\begin{equation}
\label{Ttree} T = \frac{1}{2\pi \sqrt{N} l_s} \left( 1 +
\sum_{i=3}^\infty a_i \frac{1}{N^i} \right) \ , 
\end{equation}
where we used that the leading order correction comes from the
$R^4$ term.
This seems to indicate that the Hagedorn temperature is given by
\eqref{thag} only when $N$ is large whereas for finite $N$ it has a
different $N$-dependence.
Nevertheless, we expect that there are no tree-level corrections
to the Hagedorn temperature so that \( a_i = 0 \) for all $i$.
This is easily seen to be necessary from the requirement
that \eqref{Ttree}  be equal to \eqref{thag}, since from \eqref{Ttree}
it would follow that the string tension $\hat{\tau}$
associated with Hagedorn behavior of LST
would have corrections $\hat{\tau} = N^{-1} \tau ( 1 + \Ord(N^{-1}) )$
which clearly cannot make sense as a fractional string tension
This fact also follows by noticing that the near-horizon
black NS5-brane solution is described by the
$SL(2,R)/SO(1,1) \times SU(2)$ exact CFT
\cite{Maldacena:1997cg,Sfetsos:1998wc} (see \cite{Callan:1991dj} 
for an
exact model of the extremal NS5-brane), which
implies that the metric in the string frame, and hence the temperature,
does not have tree-level corrections \cite{Tseytlin:1993df}.

Turning, to the string loop expansion, this becomes an expansion
in
$\varepsilon_L$ given in \eqref{epsD}. In particular, the
leading correction to the temperature is therefore generated by
the one-loop $R^4$ term of order
$\varepsilon_D^3 \varepsilon_L$.
Thus, using \eqref{epsD}, we can write
\begin{equation}
\label{Tcorr} T = \thg \left( 1 - \frac{\pi^2 b}{2} \frac{1}{N^2
l_s^2 u_0^2} \right) \ ,
\end{equation}
where $\thg$ is given by \eqref{thag} and the constant $b$ is a
rational number which depends on the exact form of the perturbed
solution. Since we assume that all tree-level corrections to the
temperature vanish, \eqref{Tcorr} is valid for all \( u_0 \gg
\sqrt{N} l_s^{-1} \).
Note that the one-loop correction, has generated a $u_0$ dependence
in the temperature, so that the degenerate phase-space, with constant
temperature, now expands and determines the leading order
behavior of the thermodynamics.
 On general grounds, it is
expected that the temperature $T$ in \eqref{Tcorr} is an
increasing function of the energy, which means that $b > 0$ in
\eqref{Tcorr}. Assuming $b > 0$ we see that \eqref{Tcorr} strongly
suggest that the temperature $\thg$ is a limiting Hagedorn
temperature of LST, in the sense that the temperature $T$ of the
system can come arbitrarily close to $\thg$ for higher and higher
energy $E \propto u_0^2$, but that it never can pass $\thg$.

Solving \eqref{Tcorr} for $u_0^2$ we obtain from \eqref{leadtherm}
the leading order expressions for the entropy as
a function of temperature,
\begin{equation}
\label{hagentrT}
  S(T) = \pi^3 b N \hat{V}_5
\frac{\thg}{\thg - T } \ , 
\end{equation}
with $\hat{V}_5 = V_5/(2\pi \hat{l}_s)^5 $. 
We are now in position to compare with the statistical
thermodynamics of a 5+1 dimensional closed supersymmetric string theory.
To this end we note that
from \eqref{leadtherm} and \eqref{limits} it follows that
$  \hat{l}_s E \gg ( \hat{V}_5 )^{1/5} $ so that the
five spatial world volume dimensions are effectively compact, and hence
the expected temperature form of the entropy is given by \eqref{entr0}.
It then follows that
 the entropy \eqref{hagentrT} of the one-loop corrected NS5-brane
solution and the entropy \eqref{entr0} derived by statistical mechanics
for closed strings agree in that they exhibit identical temperature
dependence.
Moreover, this
means that the constant of proportionality in \eqref{entr0} for
LST with 16 supercharges is predicted to be
\begin{equation}
\label{k} k = \pi^3 b N \hat{V}_5 \ .
\end{equation}
However, there is a remaining puzzle, which has been addressed in
\cite{Berkooz:2000mz}. While for the free string, the energy
in the compact case is intensive due to the dominance of long strings, for LST
the supergravity dual predicts an extensive energy.
 This hints at a
new universality class of interacting strings with a
 strong self-attractive potential, so that long strings are
 suppressed and the strings prefer to be coiled \cite{Berkooz:2000mz}.

\section{Hagedorn behavior of OD$p$-theories}

Recently, new non-gravitational six-dimensional theories have been
constructed by turning on a critical electric field on the NS5-brane
\cite{Gopakumar:2000ep,Harmark:2000ff}. These theories are called
Open D$p$-brane (OD$p$) theories%
\footnote{The OD1 and OD2-theories are called $(1,1)$ and $(2,0)$
Open Brane Little String Theories (OBLSTs) in \cite{Harmark:2000ff}.},
and for a given $p=0,...,5$,
they contain light open D$p$-brane excitations.
They are obtained from the D$p$-NS5 brane bound state in the limit
\begin{equation}
\label{ODplimit}
l_s \rightarrow 0 \spa
g_s = \tilde{g} \varepsilon^{\frac{3-p}{4}} \spa
\varepsilon = l_s^4 / \leff^4
\end{equation}
keeping $\tilde g$ and $\leff$ finite. 
The light open $p$-brane has tension
$\frac{1}{2} T_{ODp}$ where
\( T_{ODp} = ((2\pi)^p \tilde{g} \leff^{p+1})^{-1} \).
The theory obtained in this limit also contains
closed strings of tension $1/ (2\pi \leff^2 )$
\cite{Gopakumar:2000ep,Harmark:2000ff}.
To see that this is the correct tension, one may
compute the tension of a closed string soliton in the low-energy
SYM theory, which gives \( (2\pi)^2 / \gym^2 = 1 / (2\pi \leff^2) \)
since $\gym^2 = (2\pi)^3 \leff^2 $.
Therefore, $\leff$ is the string length associated with
``little'' closed strings.

The supergravity solution for the D$p$-NS5 brane bound state is 
\begin{eqnarray}
ds^2 &=& D^{-1/2} \Big[ D \Big( -f dt^2 + (dx^1)^2 + \cdots + (dx^p)^2 \Big)
\nn \\ &&
+ (dx^{p+1})^2 + \cdots + (dx^5)^2
+ H \Big( f^{-1} dr^2 + r^2 d\Omega_3^2 \Big) \Big] \ , 
\end{eqnarray}
\begin{equation}
e^{2\phi} = H D^{\frac{p-3}{2}} \ , 
\end{equation}
\begin{equation}
A_{01\cdots p} = (-1)^p \sin \hat{\theta} \coth \hat{\alpha}
D H^{-1} 
\spa
A_{(p+1) \cdots 5} = (-1)^p \tan \hat{\theta} H^{-1} \ , 
\end{equation}
\begin{equation}
f = 1 - \frac{r_0^2}{r^2} \spa
H = 1 + \frac{r_0^2 \sinh^2 \alpha}{r^2} \spa
D^{-1} = \cosh^2 \theta - \sinh^2 \theta H^{-1} \ ,  
\end{equation}
\begin{equation}
\sinh^2 \alpha = \cos^2 \hat{\theta} \sinh^2 \hat{\alpha} \spa
\cosh^2 \theta = \frac{1}{\cos^2 \hat{\theta}}  \ . 
\end{equation}
The near-horizon limit is given by the limit \eqref{ODplimit} along with
\begin{equation}
\cosh \theta = \frac{1}{\sqrt{\varepsilon}} \spa
r = \tilde{r} \sqrt{\varepsilon} \spa
r_0 = \tilde{r}_0 \sqrt{\varepsilon} \spa
x^i = \tilde{x}^i \sqrt{\varepsilon} \ ,\ i = p+1,...,5
\end{equation}
from which we obtain the near-horizon solution%
\footnote{The non-extremal D$p$-NS5 solution and its near-horizon limit
were given previously in \cite{Harmark:2000ff} for $p=1,2$.
The extremal D$p$-NS5 solution and its near-horizon limit 
were given in \cite{Alishahiha:2000pu}. For $p=2,3$ it was also given
in \cite{Alishahiha:1999ci}.}
\begin{eqnarray}
\label{NHsolmet}
ds^2 &=& H^{-1/2} \frac{R}{\tilde{r}} \left[ H \frac{\tilde{r}^2}{R^2}
\Big( -f dt^2 + (dx^1)^2 + \cdots + (dx^p)^2 \Big)
\right. \nn \\ && \left.
+ (d\tilde{x}^{p+1})^2 + \cdots + (d\tilde{x}^5)^2
+ H \Big( f^{-1} d\tilde{r}^2 + \tilde{r}^2 d\Omega_3^2 \Big) \right] \
, 
\end{eqnarray}
\begin{equation}
\label{strcoup}
g_s^2 e^{2\phi}
= \tilde{g}^2 \frac{\leff^2 N}{\tilde{r}^2}
\left(1+ \frac{\tilde{r}^2}{\leff^2 N} \right)^{\frac{p-1}{2}} \ , 
\end{equation}
\begin{equation}
T_{Dp} A_{01\cdots p} = (-1)^p T_{ODp} \frac{\tilde{r}^2}{\leff^2 N}
\spa
T_{D(4-p)} A_{(p+1)\cdots 5} = (-1)^p T_{OD(4-p)} H^{-1} \ , 
\end{equation}
\begin{equation}
\label{NHsolfH}
f = 1 - \frac{\tilde{r}_0^2}{\tilde{r}^2} \spa
H = 1 + \frac{\leff^2 N}{\tilde{r}^2} \ , 
\end{equation}
with \( T_{Dp} = ((2\pi)^p g_s l_s^{p+1})^{-1} \).
The curvature in units of $l_s^{-2}$ is
of order $ \mathcal{C} = \left( N^2 + \tilde{r}^2 N \leff^{-2} \right)^{-1/2}$ 
and we need \( \mathcal{C} \ll 1 \) and \( g_s e^{\phi} \ll 1 \)
in order for the near horizon solution \eqref{NHsolmet}-\eqref{NHsolfH} to 
describe OD$p$-theory.
We see that the solution \eqref{NHsolmet}-\eqref{NHsolfH} has two phases,
the NS5-brane phase for \( \tilde{r} \ll \leff \sqrt{N} \) and the
delocalized D$p$-brane phase for \( \tilde{r} \gg \leff \sqrt{N} \).
In the NS5-brane phase the theory is the ordinary $(1,1)$ LST or $(2,0)$ LST
which reduces to ${\rm SYM}_{5+1}$ or $(2,0)$ SCFT at low energies.
The D$p$-brane phase is instead a theory of
closed ``little'' strings on a special geometry.

The leading order thermodynamics is%
\footnote{This has previously been
computed for the OD1 and OD2-theories in \cite{Harmark:2000ff}.}
\begin{equation}
\label{ODtherm}
T = \frac{1}{2\pi \leff \sqrt{N}} \spa
S = \leff \sqrt{N} \frac{\tilde{V}_5}{ \tilde{g}^2 (2\pi)^4} \tilde{r}_0^2 \spa
F = 0 \spa
E = \frac{\tilde{V}_5}{(2\pi)^5} \tilde{r}_0^2 \ . 
\end{equation}
The curvature at the horizon radius in units of $l_s^{-2}$ and
the effective string coupling squared at the horizon are
respectively of order
\begin{equation}
\varepsilon_D = \left( N^2 + \tilde{r}_0^2 N \leff^{-2} \right)^{-1/2}
\sp \varepsilon_L = \tilde{g}^2 \frac{\leff^2 N }{\tilde{r}_0^2}
\left(1+ \frac{\tilde{r}_0^2}{\leff^2 N} \right)^{\frac{p-1}{2}} \ . 
\end{equation}
The leading order thermodynamics \eqref{ODtherm}
is valid for \( \varepsilon_D \ll 1 \)
and \( \varepsilon_L \ll 1 \).
As for ordinary LST, the thermodynamics \eqref{ODtherm} exhibits leading
order Hagedorn behavior with Hagedorn temperature
\begin{equation}
\label{ODtemp}
\thg = (2 \pi \leff \sqrt{N})^{-1} \ . 
\end{equation}
We see that this temperature is the Hagedorn temperature associated
with the ``little'' closed strings in OD$p$-theory \cite{Harmark:2000ff}.
In fact, when the supergravity dual \eqref{NHsolmet}-\eqref{NHsolfH}
is valid we have
that $T \sim \thg$, which we take to mean that the ``little'' closed
strings are the fundamental degrees of freedom in this region of phase
space.
We note that the Hagedorn temperature \eqref{ODtemp}  was also found
using different arguments in \cite{Gubser:2000mf}.

We now want to compute the leading correction to the temperature,
as done above for ordinary LST, since this gives the temperature dependence
of the entropy.
Since we want to consider the high energy behavior of the thermodynamics
we assume that we are in the D$p$-brane phase $\tilde{r}_0 \gg \leff \sqrt{N}$.
Contrary to the NS5-brane case, the leading correction in this
case comes from the leading tree-level $R^4$ term at order $\varepsilon_D^3$.
We therefore find
\begin{equation}
T = \thg \left( 1 - a^3 \frac{\leff^3}{N^{3/2} \tilde{r}_0^3} \right)  \ , 
\end{equation}
where $a$ is an undetermined constant.
The resulting leading order expression for the
entropy as a function of temperature is
\begin{equation}
\label{correntr}
S(T) = a^2 \frac{\leff^3}{\sqrt{N}} \frac{\tilde{V}_5}{\tilde{g}^2 (2\pi)^4}
\left( \frac{\thg}{\thg - T} \right)^{2/3} \ . 
\end{equation}
Comparing \eqref{correntr} with the entropy \eqref{hagentrT} for ordinary LST
we see that the critical behavior for high energies is different.
For ordinary LST the entropy is proportional to $(\thg-T)^{-1}$,
while for the OD$p$-theories the entropy is proportional to $(\thg-T)^{-2/3}$,
as already noticed in \cite{Harmark:2000hw,Harmark:2000wv,Harmark:2000ff}.
Even though the high energy behavior
of OD$p$-theory is dominated by ``little'' closed strings just like
for ordinary LST, we interpret this difference as arising from
the fact that the ``little'' closed strings of OD$p$-theory live
 on a different kind of geometry, e.g. for $p=1$ the space-time is
 space-time non-commutative.
We call this phase the deformed LST phase.
It would be very interesting to reproduce the critical exponent
$-\frac{2}{3}$ of the entropy
from the statistical mechanics of closed strings
on one of the OD$p$-theory geometries.


\vskip .5cm 
\noindent {\bf Acknowledgments.} We thank the organizers for
an interesting and pleasant conference.


\section*{References} 



\providecommand{\href}[2]{#2}\begingroup\raggedright\endgroup
\end{document}